\documentstyle [12pt]{article}
\input epsf.tex
\begin{document}

\title{Systematic effects in the determination of the $\pi~N\!~N$ Coupling
from $\bar p p \rightarrow \bar n n$ differential cross section}
\author{
Torleif Ericson\thanks{Also at
Theory Division, CERN,
CH-1211 Geneva 23, Switzerland} \\
{\small The Svedberg Laboratory, Uppsala University, Box 533,
S-75121 Uppsala, Sweden }\\ 
 \and {\small and}\\
\and Beno\^\i t Loiseau   \\
{\small Division de Physique Th\'{e}orique\thanks{Unit\'{e} de Recherche des Universit\'{e}s
 Paris 11 et Paris 6 associ\'{e}e au CNRS}, Institut de Physique
Nucl\'{e}aire,
F-91406  Orsay Cedex} \and {\small and LPTPE
Universit\'e Pierre et Marie Curie, F-75252 Paris Cedex 05,  France} \\ }
\date{October 8 1996}
\maketitle
{\center
\begin {abstract}
We show that the $\pi NN$ coupling constant extracted model- independently
from $\bar pp$ charge exchange is subject to a systematic correction, and,
more importantly, that the strong absorption in the critical region prevents
a determination of the coupling constant to high precision using this
process. This attenuates the possible conflict with the value determined
from the $np$ charge exchange cross sections. \end {abstract} }

The value of the $\pi$NN coupling constant has become a topic of hot debate
in the last few years \cite{ERI93}.  In addition from the 'classical'
determinations from $\pi$N scattering and forward NN dispersion relations
with a 'high' value of $g^2$ near 14.3, the Nijmegen group has argued a
lower value of about 13.6 on the basis of their extensive analysis of NN
data using the Nijmegen model \cite{KLO91,STO93}. The issue of its precise
value is an important question not only because it is a fundamental constant
to nuclear physics \cite{ERI95}, but also because it is of great importance
also for the understanding of chiral symmetry breaking.  Thus, the
experimental error in the pion-nucleon coupling constant is the
principal obstacle in the precise testing of the validity of the
Goldberger-Treiman relation as predicted from chiral symmetry
breaking \cite{FUC90,DOM85}.  This relation would give $g^2=12.81$, if it
were exact, which is not expected, however.  To resolve the problem
unambiguously we have undertaken a detailed study of the possibility of
using methods of extrapolation to the pion pole in NN scattering for an
accurate and model independent determination of the coupling constant
directly from data.  A first report on this work, based on high precision
absolutely normalized data in the pole dominated region of np elastic charge
exchange at 162 MeV, points to a high value of $g^2=14.60\pm0.30$
\cite{ERI95}. The dominant source of uncertainty is the experimental
overall normalization.  New, as yet unpublished, additional data give a
slightly lower and more accurate value $g^2=14.4\pm0.2$ \cite{OLS96}\cite {ERI96}. The
pole extrapolation method was originally suggested by Chew using a
polynomial expansion \cite{CHE58}.  Its practical use is fraught with
problems associated with systematic errors and instabilities which must be
controlled and well understood.  We have tentatively solved this
problem on the one hand by stabilizing the extrapolation procedure using
appropriate comparison functions, on the other hand by exploring
systematical errors analyzing large numbers of 'pseudo experiments' generated
from models with known coupling constants.  This program is still being
pursued on a wider body of $np$ charge exchange data.

 The corresponding information is in principle also available from the 
$\bar p p\rightarrow \bar n n$ charge exchange reaction, which has the same pole
structure. Recently precise data on this reaction was obtained in the
experiment PS206 by Birsa et al. \cite{BIR94} at 601 MeV/c (= 176 MeV kinetic energy) and the
quality of these data are comparable and even superior to those of the best
data on $np$ charge exchange in a wide range of similar energies
\cite{BRA94a}.  A preliminary analysis 
 demonstrated strikingly that indeed both $p\bar p$ and $np$ charge exchange
data extrapolate approximately to the same pole residue \cite{BRA94a}.  However, the
further analysis \cite{BRA94b} based on the classical Chew method  gave the very low value
$f_c^2=0.0708\pm0.0016\pm 0.0011$ ($g^2=12.80\pm0.29\pm 0.20$) (statistical and
normalization errors, respectively) , about 10\% below our value above and also
well below
 the
Nijmegen value. This value is a cause of concern and raises the
question of consistency with $np$ charge exchange and even more seriously of the
overall validity of the approach as such. The analysis did not attempt to analyze and evaluate the
systematic errors in the procedure. The  present note aims to elucidate
these questions as well as 
 to clarify the structure
of the contributions to antiproton charge exchange and their relations to the
pion pole terms.

We first examine the contributions to the unpolarized differential cross
section so as to illustrate the information that must be described by any
extrapolation procedure. Here the total amplitudes and cross sections are
defined in the usual fashion in terms of the five amplitudes $a,b,c,d,e$
allowed by the invariance properties \cite{BYS78,LAF92}. Both the CM unpolarized
cross section as well as the polarization transfer one are incoherent
combinations of 5 amplitudes:
$$\frac {\rm d\sigma}{\rm d\Omega}(t)=1/2(|a|^2+|b|^2+|c|^2+|d|^2+|e|^2);$$
\begin{equation}
\frac {\rm d\sigma}{\rm d\Omega}(t)~(1-K_{n00n})=|c|^2+|d|^2,
\end{equation}
where $t$ is the squared 4-momentum transfer from the proton to the neutron.
 Taking into account the difference between the proton $(M_{p})$ and neutron
$(M_{n})$ masses one has in the centre-of- mass
 system for the proton momentum k and scattering angle $\theta$:
\begin{equation}
 t=+2k\sqrt{k^2-M_n^2+M_p^2}\cos\theta -2k^2+M_n^2-M_p^2.
\end{equation}

There are five regularized pion Born amplitudes
 and in this case
the r-space $\delta$-function has been subtracted  \cite{GIB94}~\cite {massdif}:
$$a_{\pi}+b_{\pi}=0; $$
$$a_{\pi}-b_{\pi}=F(t); $$
$$c_{\pi}+d_{\pi}=F(t)(1-3\Pi(t));$$
$$c_{\pi}-d_{\pi}=F(t);$$
\begin{equation}
e_{\pi}=0,
\label{ope}\end{equation}
where
\begin{equation}
F(t)\equiv \frac{g^2}{\sqrt{s}}
\left(\frac{{\Lambda}^2 - m_{\pi }^2}{{\Lambda}^2- t}\right)^2;\
\Pi(t)\equiv \frac{ t}{ t- m_{\pi }^2}.
\label{ffp}\end{equation}
In Eq. (4) the charged pion mass is denoted by $m_{\pi}$ and the range of the form
factor is chosen to be $\Lambda=800$ MeV. This corresponds to an rms radius of .6 fm for the
nucleonic pion source.

 It is in
practice convenient to chose a representation for the amplitudes such that the pion pole term appears
in one single term, namely $(c+d)/\sqrt {2}$ for the $p\bar p$ case and $(b-d)/\sqrt {2}$ for the $np$
one.  This regroups the pole term completely into the real part of these amplitudes, respectively, 
 and they combine to an exact square in the cross section. 
These are therefore crucial to the extrapolation and they are also the dominant terms in the physical
region.  All the remainder is represented by terms which slowly and smoothly decrease with increasing
$-t$.  The
differential cross section and polarization transfer then become
$$
\frac {\rm d\sigma}{\rm d\Omega}(t)
=1/4[|a+b|^2+|a-b|^2+|c+d|^2+|c-d|^2+2|e|^2)];$$
\begin{equation}
\frac {\rm d\sigma}{\rm d\Omega}(t)~(1-K_{n00n})
=\frac{1}{2}[|c+d|^2+|c-d|^2]
\end{equation}

Let us now examine the behavior of the contributions to the cross section and their variation
with $t$ in more detail. Such a study is most
readily done using a model which represents the main physics even if it only qualitatively
reproduces the data.  The Paris antiproton model is sufficient for this purpose and readily
reproducible \cite{PIG94}. The characteristic shape and magnitude of the terms in eq. (5) are
illustrated for the pion Born terms of eqs. (3-4) and for the Paris model in
Figs. 1a and 1b.  Not unexpectedly, absorption  is a prominent feature for
antiproton charge exchange.  Compared to $np$ charge exchange the cross
section is  quenched by a factor 4-5 for momentum transfers of a few pion
masses, the  region most sensitive to the pion pole.   This feature is
strikingly apparent also in the comparison of the $np$ and the  $p\bar p$
pole extrapolation in fig. 2.  There is  an important difference in
the physics between the two cases. In the $np$ case the amplitudes have
imaginary parts only from unitarity. They are mostly nearly real and there
are in practice only two important non-pole terms.  In the antiproton charge
exchange real and imaginary non-pole amplitudes are both important and four
such terms contribute substantially.  Fortunately our method permits us to
ignore the detailed dynamics of these terms.  In both case the 
 term  containing the  pole amplitude has a zero 
 at $t= -m^2_{\pi}/2$ in the Born approximation independent of form
factors.  The corresponding  minimum survives with a slight shift in the
model cross sections as seen in Figs. 1a and 1b.  Although the value is
quenched in the antiproton case this suggests that similar extrapolation
methods can, at least in principle, be  used in both cases. Practice
appears to confirm this.  The minimum is accompanied by a steep rise of the
 differential cross section at
small angles.  It is this feature that is responsible for the overall minimum in
the cross section, which is partly masked by slowly varying background terms. In
the case of $np$ scattering this minimum transforms into a shoulder in the cross
section, but the corresponding structure is also present in that case.

We note in passing that polarization transfer in principle allow a clean
 separation of the terms
containing the pion pole contributions from other terms as is apparent 
from eq. (5).  However, the
use of this possibility still requires the accurate knowledge of the
 unpolarized cross sections
for a direct extrapolation to the pion pole. Since the eliminated terms are 
not dominant, even a
perfect knowledge of the  polarization is only a minor constraint on the 
pion coupling constant.
 The
polarization information is of no particular use in the present context.

 We have first followed a  procedure identical to that of Bradamante et al.,
analyzing their data using the Chew method \cite {BRA94a}, \cite{BRA94b}. In this case one
defines the function y(x), which extrapolates the experimental data
smoothly to the pion pole:
\begin{equation}
y(x)= \frac {sx^2}{m_{\pi}^4g_R^4} \frac {{ \rm d}  \sigma}{{\rm d} \Omega}(x)
= \sum _{{ \rm i=0}}^{{ \rm n-1}}a_ix^i. \label{eq: CHEW}
\end{equation}
Here $s$ is the square of the total energy and $x= -t+m_{\pi}^2$. We will use the
charged pion mass $m_{\pi }\equiv 1$  as the mass scale in the discussion.
At the pion pole $x=0$ the Chew function gives
\begin{equation}
y(0) \equiv a_0 \equiv N g^4/g_R^4,
\label{eq:poleCHEW}
\end{equation}
where N is the normalization of the experimental data, which may differ from
the true value. The model independent determination of the coupling constant $g^2$ requires
accurate single-energy data with absolute normalization N of the unpolarized
differential cross section. The error in the coupling constant determined by such extrapolation
methods  is always proportional to $\sqrt N$. 

We accurately reproduce the Bradamante analysis for the $\bar
pp$ reaction  using the Chew method and the results are given in Table 1.   However, we find minor
discrepancies in their simultaneous analysis of the Uppsala 96 MeV $np$ scattering data, which they
use as a comparison.  We traced this to their arbitrary omission of three data points. 
 Therefore, at this first step
our analysis reproduces  their  low value of the coupling constant.

The next problem is that of systematical errors in the extrapolation
procedure and true extrapolation errors.  Here the $\bar pp$ case differs
importantly from the $np$, since absorption is a major feature.  Examination of the
Chew function in fig. 2 shows that it has a pronounced minimum near $0^{\circ} $,
the physical point closest to the value to be determined.  The extrapolation point
is about a factor of 4 larger than this value and lies opposite to the main
trend of the function to be extrapolated.  As a matter of principle this is a
particularly unfavorable situation in any accurate extrapolation procedure. Another way of
stating this problem is to note that the absorption produces an important reduction of the pion
pole effects in the physical region, which is to say that it reduces the sensitivity to the
coupling constant which we want to determine.  Even so there remains a substantial
sensitivity on the pion-nucleon coupling constant as already  demonstrated by Bradamante
et al. \cite{BRA94a,BRA94b}.  The accuracy to which it can be determined is however a more
delicate and quantitative question.

 To investigate this matter we first use   models with known $g^2$ for which we can generate
randomly 'data points' at the exact angles and with the same statistical errors as in the actual
experiment.  For this we generate 10.000 equivalent pseudo experiments which we analyze in the same
way as the actual experiment.  As models we use a solution of a preliminary
analysis of the Nijmegen group  \cite{TIM95} and the Paris potential
prediction  \cite{PIG94}. However any other model with known pion exchange
 would do as well for the present purpose if it approximately reproduces the data.
From  Table 1 we  find that the Chew extrapolation approach
requires 6 polynomial terms to describe these 'experimental' data appropriately.
 From the results we conclude that the Chew
extrapolation procedure systematically underpredicts the coupling constant by
2\% ($\delta g^2\simeq 0.3$) for the case of 47 data points (n=6).  This shift would appear even if the precision of the data
were higher than at present. It is thus necessary to correct for such  systematic shifts in the
determination of the coupling constant using the Chew method.

However,  very little
information on the pion pole is contained in the region of large momentum transfers.  We  therefore
also truncated the data at $x_{max}=4.9m_{\pi}^2$ (the 30 first points), so as to be able to use fewer
parameters in the extrapolation.  One easily persuades oneself that the tensor
amplitude makes it physically unreasonable to expect a description of the data with a
polynomial of less than 5 terms in the original Chew polynomial extrapolation
procedure and that data with large momentum transfer call for at least 6 terms.

 In the case of 30 points, the analysis of the pseudodata for
both the Paris and Nijmegen models in Table 2 demonstrates that 5 terms in the polynomial give a
perfect fit to data such as these, although with a systematic downward shift of $g^2$ by 0.3 (Paris)
to 0.5 (Nijmegen).   The extrapolated
value is very low (12.26) and it is not more than about $12.8\pm 0.5$, even when corrected for the
systematic shift. The shift disappears with one additional term in the polynomial, but at the cost of
a  much larger extrapolation  error.
 For the corresponding
experimental 30 data points 
the  $\chi^2/DoF$ is only 0.46, which is
  an unexpectedly low value statistically.  It reflects partly that the errors given by PS 206
consist of statistical and systematic errors added in quadrature \cite {BIR94}.
 Corrected for
systematic shifts the Chew model in this case gives $12.84\pm.46$ for $n=5$ and $ 13.24 \pm 1.15$
for $n=6$. 
 The low $\chi^2$ raises the possibility of some correlations in the 
systematic errors, in particular at low momentum transfers. In every case the
systematic  error of 1.5\%, i.e. $\pm .2$, due to the normalization uncertainty must be added to the
overall uncertainty.
These results are compatible with the Chew analysis in the larger data range with 47 data
points. The latter appears to be more accurate with a value $g^2~=~13.1\pm .3$ as seen in
Table 1 for $n=6$. This is somewhat disconcerting since the additional region is insensitive
to the pion pole and the polynomial expansion is questionable over such a large region.

We now turn to the question of the precision that can be attained reliably in the
analysis, since the errors given above are only the formal statistical uncertainty of the Chew
method.  To investigate this point it is convenient to use  the Ashmore method as 
described in ref. \cite{ERI95}. We expect this method to work, since both  the background terms and
the interference term  vary similarly to the $np$ case as already discussed above. It is thus
the systematics of this method that will determine the precision which can be achieved.
The Ashmore method parametrizes the differential cross
section in terms of the regularized pion Born amplitudes, but has in addition a contribution
simulating a $\rho$ meson pole with adjustable strength and shape, described with a form factor
and polynomial terms in $x$  \cite {ERI95}.  This method permits a phenomenological simulation of the
absorption effects, but the dynamical description as such is not realistic.  The virtue of the method
is its ease of application and that it permits the explicit exploration of the influence of changes in
the pion-nucleon coupling constant.  
 We  note from Table 2 that the Paris and Nijmegen model
pseudodata are described in the Ashmore model for $n$=5  with $\chi^2 / DoF$=1.00 with a moderate systematic
 shift of 0.21 and 0.29, respectively, for 30 data points.  An additional parameter as
used with $n=6$ will thus overparametrize the analysis.  

In order to test the uncertainty of the
coupling constant determined from the {\it full} range of data in the Ashmore model we fixed the
coupling constant for $n=6$ to 13.21, 14.10 and 14.43 and determined the corresponding $\chi^2$
values 40.2, 51.0 and 58.6 for the 47 data points (41 degrees of freedom).  This
should be compared to the corresponding Nijmegen description for $g^2=13.23$, which is 48.6.
Although the value of 13.23 gives a better $\chi^2$ in agreement with the other methods all of these
values give a good representation of the data as is apparent from Fig. 3. We note in particular that
the first few points near $t=0$ have a large contribution to the $\chi ^2$.  They
play thus an essential role in the discussion.  For example, one half of the difference between the
$\chi ^2 $ of the Nijmegen model with a coupling constant of 13.23 and the Ashmore model with 14.43
comes from the two points nearest $t=0$.  Since these points are also the ones with the largest
systematic errors \cite {BIR94}, they are an important potential source of systematic error in the
extrapolation. If they are omitted, the value of $g^2$ increases by $.1$ to $.2$.
Interestingly, all of the Ashmore curves have a generically similar shape which differs
systematically from the one in the Nijmegen description on the level of a few~\%.  These
observations suggest that the value of $g^2$ deduced from antiproton charge exchange cannot
be considered to fully model independent at a precision higher than 0.5 to 0.7 units (3 to
5\%).

In the case of $np$ charge exchange, which we previously investigated, it was possible to
improve the procedure considerably using a Difference Method \cite {ERI95}. This method relies on an
extrapolation of the {\it difference} of the Chew function for data and a model with a known
coupling constant.  While this method is effective in the case of $np$ charge exchange, it does not
improve significantly on the previous methods in the present case. The reason is
the scarcity of high precision
 information on the $p\bar p$ interaction other than charge exchange. This has the consequence that
either the model relies heavily on the present data on charge exchange or lacks sufficient precision
in the description of higher momentum transfer to be useful. In the former case the method would as
expected accurately reproduce the $g^2$  of the model with considerable precision, but the argument
would be circular. The method has the additional advantage of visually bringing out the details of
the extrapolation in the low $t$ region at the level of precision under discussion here. It is
therefore instructive to apply the method to  the Paris model, which has qualitative agreement with
data and incorporates constraints at large momentum transfer. 

 The results of the extrapolation of
the difference function $y_M(x)-y(x)$ versus $x$ are given in Table 2 for 30 data points and are
displayed in figure 4.  We first calibrated the method using pseudodata from the Nijmegen model. 
These demonstrate that such data can be perfectly described by $n=5$ with $\chi^2/DoF$ = 1.00 and
that the systematic  extrapolation error is then negligible. However, applied to the actual data we
find no improvement in the accuracy to which the coupling is determined.  On the other hand  Fig. 4
demonstrates that a major background has been removed. It is now easy to visualize  the consequences
of different values for the number of polynomial terms $n$ in the extrapolation. A good description
is obtained already for $n=4$ but with a substantial systematic correction. The values deduced for
the coupling constant after the correction for the systematic shifts are in this case $13.10 \pm
.16$, $12.65 \pm .45$ and $13.24 \pm 1.15$ for $n=4,~5$ and $6$, respectively. These
extrapolations give similar   results to the previous ones and they are mutually consistent. We have
also applied the method to the full set of data. In this case ($n=6$) is needed, but the conclusions
do not change.

 We have already remarked that the absolute
normalization of the cross section is a
 crucial number for the extraction of the coupling constant. Let us now discuss the uncertainties from
this source, which is independent of the errors arising from the extrapolation
uncertainty.  The Nijmegen group has achieved a good description of the present data with 
$\chi^2/(data) $ = $1.035$ with an integrated charge exchange cross section 
$\sigma_{tot}=12.14$ mb. The normalization error is most
likely the experimental uncertainty in the experiment PS206 in view of the
close fit to it.  For the Paris model the corresponding $\sigma_{tot}=13.45 $ mb.
This means  automatically  that if this normalization were used to normalize the same
experimental data, then the deduced coupling constant would increase by 5.3\%, that is
$g^2$ would increase by 0.7.  This emphasizes the importance of this question.

 Some
additional information on this point comes from other experiments.  R.P. Hamilton et al.
 \cite{HAM80} have performed a dedicated absolute measurement of the integrated
 charge exchange cross
section with energy in small steps.  By interpolation between  596 and 608 MeV/c this corresponds to
a value of  $\sigma_{tot}=11.80(11)$ at 601 MeV/c.  The systematic error is stated to be less than 3
to 5 \%, depending on the energy region.  Using 4\% as a reasonable estimate at the
present energy this means a systematic uncertainty of about 0.47 mb, completely compatible with both
the experiment PS 206 \cite {BIR94} and with the Nijmegen description.
Experimental
information can also be found in  Nakamura et al.  \cite{NAK84}.  In this case
the emphasis was on the angular shape and the forward interference dip and not on normalization.
Their integrated cross sections are larger than in the other experiments by 15 to 20\%, but the
systematic uncertainty was also larger (8.5\%). In view of the circumstances we, as the authors, do
not believe these larger values to be significant. In conclusion, there appears to be no obvious
reasons to question the presently quoted normalization in the experiment PS 206, but we emphasize
that this issue must be kept in mind as a potential source of problems.

We have here critically examined the accuracy to which the $\pi NN$ coupling constant can be extracted
from the recent precision data on $p\bar p$ charge exchange and the importance of systematic
theoretical corrections to  pole extrapolation procedures. Our conclusion is that the situation is
less favorable than for the corresponding $np$ charge exchange reaction.  This is due to the prominent
role of absorption which reduces sensitivity to the pion in the most critical region of momentum
transfers.  In particular, it is not profitable to apply the Difference Method in the present
context.  Here, the accuracy does not increase and it was this method that was the key to high
precision for $np$ charge exchange.  In spite of this limitation and the more important role of
systematics in the extrapolation procedure it is possible to extract the coupling constant to a  good
degree of precision, though with an errors of about 4 to 5\%.  Depending on the detailed
procedure we find a range of plausible deduced values from 12.8 to 13.2 from the data.  The
statistical errors are typically of order $.45$ using only the range of data sensitive to the pion
information.  Formally, a higher statistical precision
is achieved using a larger range of data.  However, we found in model studies that a nearly
indistinguishable description of the data is achieved with values for the coupling as large as
$14.1$. We therefore believe that caution should be used in quoting the formal errors above. We
recommend a value $g^2=13.0\pm .7$ with an additional systematic error of 0.3 from the
overall cross section normalization. The value that is extracted in this way is a direct
determination the coupling constant for the $\pi N\bar N$ system. This value is low, but in
view of the uncertainties it is compatible with the value for the coupling constant deduced
from other sources. 
 There is no obvious discrepancy with the value deduced from $np$ charge exchange.

 We are indebted to F. Bradamante and A. Martin
for details of the analysis of the antiproton data PS 206, 
M. Lacombe for the
solutions to the Paris antiproton potential, J. Blomgren and N. Olsson for discussions concerning on
 $np $ charge exchange data  and to R. Timmermans for information on 
the results of the Nijmegen
 analysis of the antiproton charge exchange data PS 206.

\begin{table}
\caption {Results applying the Chew method to the Birsa et al.
data points at 176 MeV [10] and to the corresponding Nijmegen and Paris
antinucleon  pseudo-data (see text). The model coupling constants are
$g^2_{Nijmegen}=13.23$ and $g^2_{Paris}=14.43$ with $\delta g^2$ the systematic shift
from the true model value.} 
\begin{center}
\label{Tab. Chew}
\medskip
\begin{tabular}{|c||ll||llr||llr|}
\hline
\multicolumn {9}{|c|}{Chew Method}\\
\hline
 &\multicolumn{2}{c||}{PS206}
 &\multicolumn{3}{c||}{'Nijmegen'}
 &\multicolumn{3}{c|}{'Paris'} \\
 \hline
n&$\chi ^ 2/DoF  $&\multicolumn{1}{c||}{$ g ^ 2 $}
 &$\chi ^ 2/DoF  $&\multicolumn{1}{c}{$  g ^ 2  $}&$  \delta g ^ 2  $
 &$\chi ^ 2/DoF  $&\multicolumn{1}{c}{$  g ^ 2  $}&$  \delta g ^ 2 $\\
 \hline
\ignorespaces
5& 1.28  &$      11.78 \pm 0.15 $
 & 1.77  &$      11.53 \pm 0.16 $& ~1.70
 & 2.54  &$      12.42 \pm 0.14 $& ~2.01\\
6& 0.90  &$      12.76 \pm 0.27 $
 & 1.00  &$      12.90 \pm 0.27 $& ~0.33
 & 1.01  &$      14.17 \pm 0.24 $& ~0.26\\
7& 0.79  &$      11.67 \pm 0.58 $
 & 1.00  &$      13.10 \pm 0.51 $& ~0.13
 & 1.00  &$      14.47 \pm 0.45 $& -0.03 \\
\hline \hline
\end{tabular}
\end{center}
\end{table}

\begin{table}
\begin{center}
\caption {Results applying different extrapolation methods to the
30 first data points of PS 206 [10] as well as to the
corresponding Nijmegen and Paris model pseudodata.} 
 \label{Tab. Chew2} \medskip
\begin{tabular}{|c||ll||llr||llr|} \hline
\multicolumn {9}{|c|}{Chew Method}\\
\hline
 &\multicolumn{2}{c||}{PS206}
 &\multicolumn{3}{c||}{'Nijmegen'}
 &\multicolumn{3}{c|}{'Paris'} \\
 \hline
n&$\chi ^ 2/DoF  $&\multicolumn{1}{c||}{$ g ^ 2 $}
 &$\chi ^ 2/DoF  $&\multicolumn{1}{c}{$  g ^ 2  $}&$  \delta g ^ 2  $
 &$\chi ^ 2/DoF  $&\multicolumn{1}{c}{$  g ^ 2  $}&$  \delta g ^ 2 $\\
 \hline
\ignorespaces
5& 0.47  &$      12.26 \pm 0.46 $
 & 1.00  &$      12.85 \pm 0.44 $& ~0.48
 & 1.00  &$      14.11 \pm 0.39 $& ~0.32\\
6& 0.46  &$      13.12 \pm 1.15 $
 & 1.00  &$      13.11 \pm 1.17 $& ~0.12
 & 1.00  &$      14.38 \pm 1.03 $& ~0.05\\
\hline \hline
\multicolumn {9}{|c|}{Ashmore Method}\\
\hline
n&$\chi ^ 2/DoF $&\multicolumn{1}{c||}{$ g ^ 2 $}
 &$\chi ^ 2/DoF $&\multicolumn{1}{c}{$ g ^ 2 $}&$ \delta g ^ 2 $
 &$\chi ^ 2/DoF $&\multicolumn{1}{c}{$ g ^ 2 $}&$ \delta g ^ 2 $ \\
 \hline
4& 0.44  &$ 12.40 \pm 0.23 $
 & 1.04  &$ 12.70 \pm 0.22 $& ~0.53
 & 1.08  &$ 13.90 \pm 0.20 $& ~0.53\\
5& 0.46  &$ 12.43 \pm 0.34 $
 & 1.00  &$ 12.94 \pm 0.33 $& ~0.29
 & 1.00  &$ 14.22 \pm 0.30 $& ~0.21\\
6& 0.46  &$ 13.43 \pm 1.55 $
 & 1.00  &$ 13.14 \pm 1.61 $& ~0.09
 & 1.00  &$ 14.38 \pm 1.43 $& ~0.04\\
\hline 
\end{tabular}
\begin{tabular}{|l||ll||lrll|} \hline
 \multicolumn {6}{|c}{Difference
Method}&\\
 \hline
&
\multicolumn{2}{c||}{Paris-PS206 }&
\multicolumn{3}{c}{Paris-'Nijmegen'}&\\
\hline
n  
 &$  \chi ^ 2/DoF  $&\multicolumn{1}{c||}{$  g ^ 2  $}
 &$  \chi ^ 2/DoF  $&\multicolumn{1}{c}{$  g ^ 2  $}&$ \delta g ^ 2 $&\\
\hline
3
 & ~4.71 &$   15.13  \pm 0.06 $&
  ~3.67 &$   14.98  \pm 0.06 $& -1.75&\\
4 
 & ~0.70 &$   13.69  \pm 0.16 $&
  ~1.10 &$   13.82  \pm 0.16 $& -0.59&\\
5
 & ~0.45 &$   12.62  \pm 0.45 $&
  ~1.00 &$   13.20  \pm 0.43 $& ~0.03&\\
6
 & ~0.46 &$   13.03  \pm 1.15 $&
  ~1.00 &$   13.02  \pm 1.17 $& ~0.21&\\
\hline \hline
\end{tabular}
\end{center}
\end{table}

\begin{figure}[htb]
\hbox{\epsfxsize=12.truecm \epsfbox{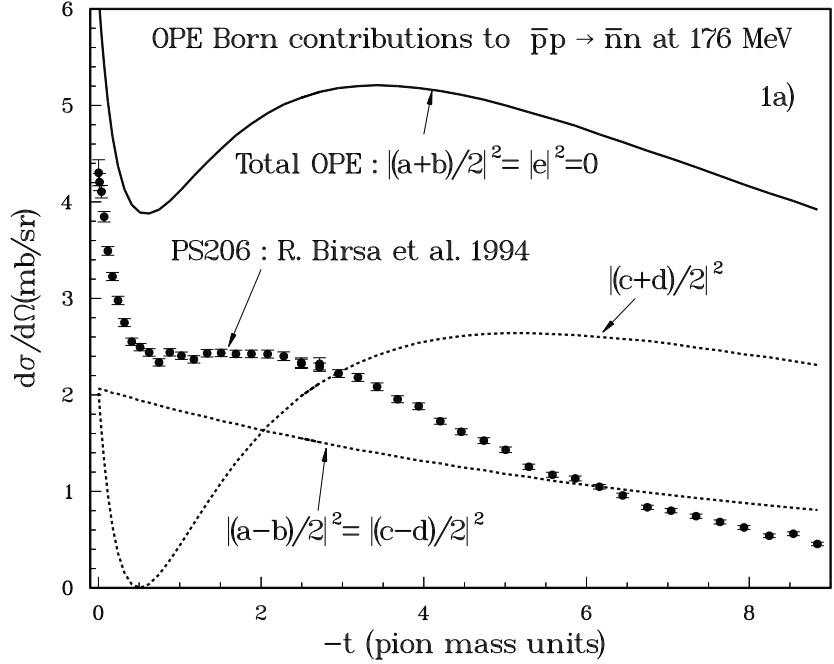}}
\vspace{9pt}
\hbox{\epsfxsize=12.truecm \epsfbox{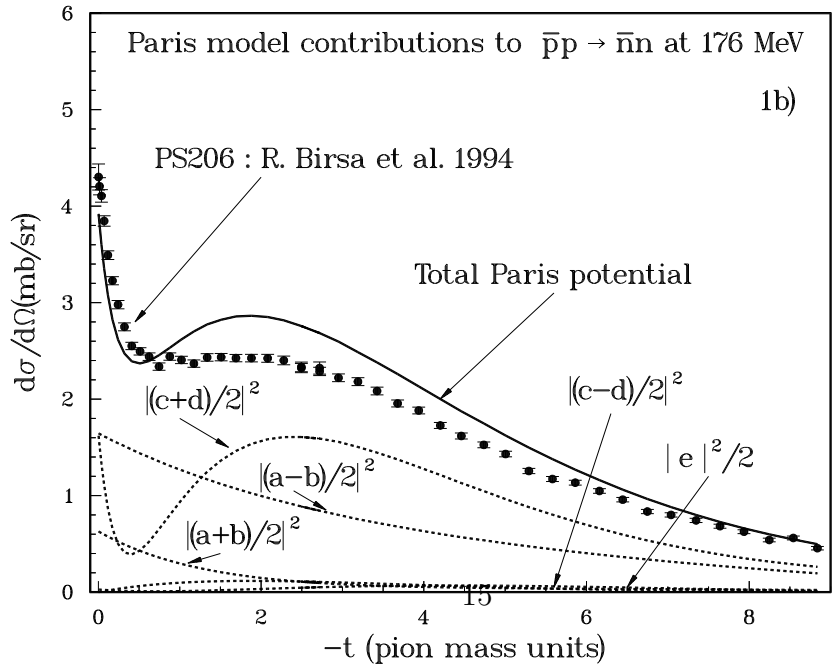}}
\caption {The magnitude and structure of the different contributions in eq. 5 to the
differential $\bar pp\rightarrow \bar  nn$ charge
exchange cross section  at 176 MeV: a) using the pion Born amplitudes of
eqs.  3 and 4;  b) using the Paris antinucleon model. 
The pion pole contributions appear in  $|c+d|^2$ only.}
\label{fig 1 a,b}
\end{figure}

\begin{figure}
\vspace{9pt}
\hbox{\epsfxsize=14.truecm \epsfbox{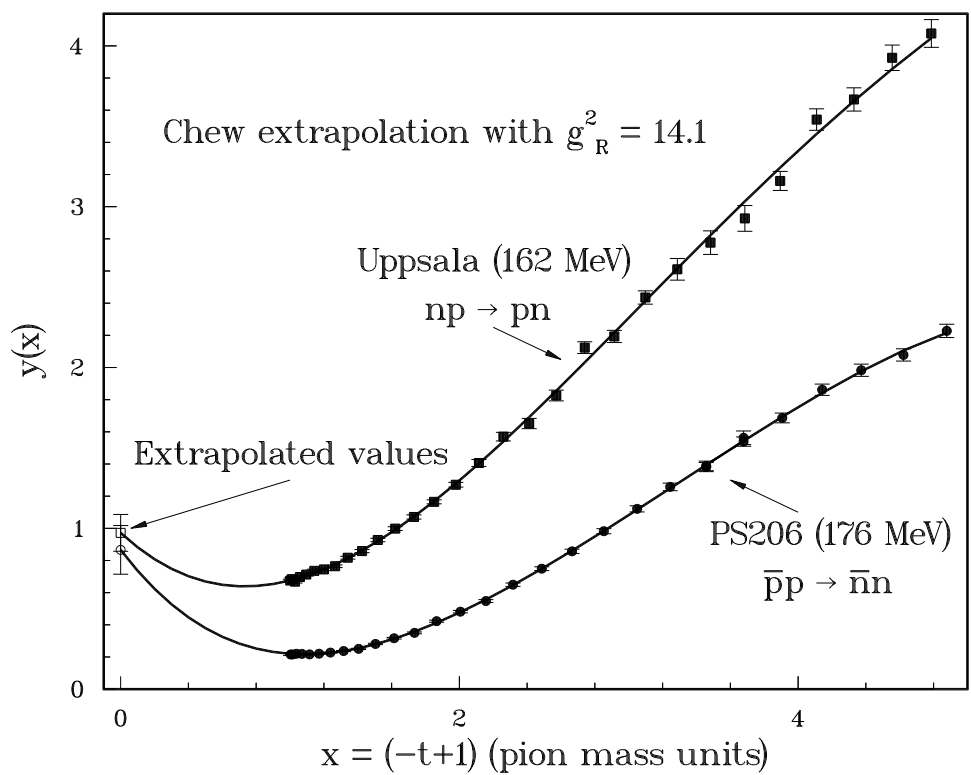}}
\caption {Comparison of the Chew pole extrapolation for $np$ and $\bar pp$ charge
exchange at 162 MeV [4] and 176 MeV [10], respectively. The
reference coupling constant is 14.1.  For details, see text. } \label{fig 2}
\end{figure}

\begin{figure}
\vspace{9pt}
\hbox{\epsfxsize=14.truecm \epsfbox{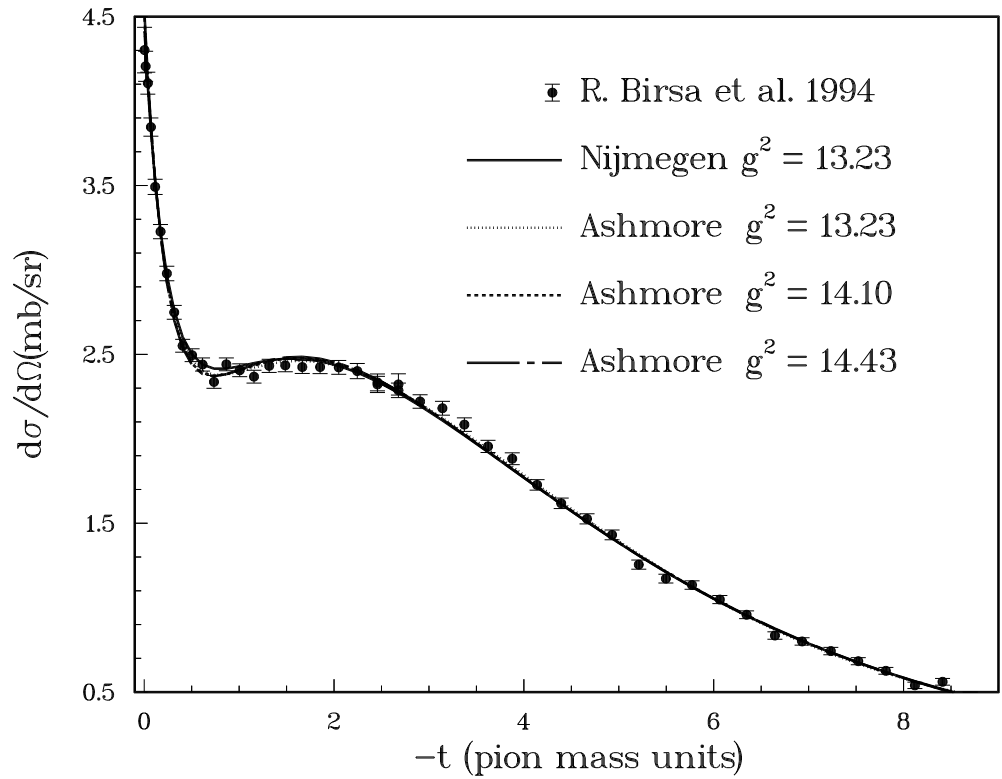}}
\caption { Fit to the data of Birsa et al. [10]  by the Nijmegen group ($g^2_{\pi NN}=13.23$) [18]
and using for the $n=6$ Ashmore model with 3 different values for the $\pi NN$ coupling ($g^2=13.23$,
$14.10$ and $14.43$, included in the definition of $n=6$). Note the similar quality of the
descriptions.}\label{fig 3} \end{figure}

\begin{figure}
\vspace{9pt}
\hbox{\epsfxsize=14.truecm \epsfbox{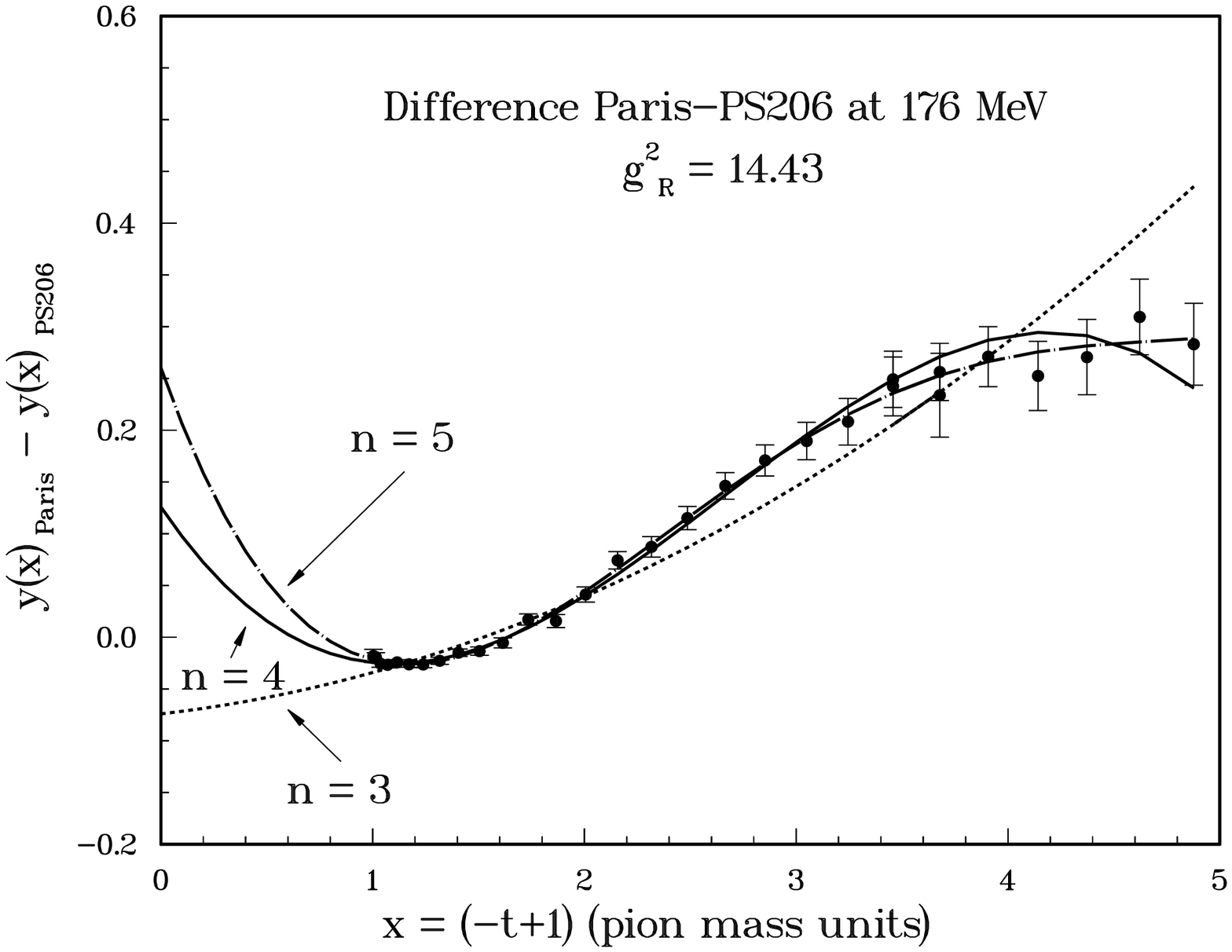}}
\caption {Extrapolation to the pole of the difference between the Chew function for
the Paris model [17] and for PS206 [10] at 176 MeV for n=3, 4 and
5. Note that n = 4 or 5 is required by the data.} \label{fig 4} \end{figure}

\begin{thebibliography}{99}
\bibitem{ERI93}
For a review, see T.E.O. Ericson, Nucl. Phys. {\bf A543}, 409c (1993).

\bibitem{KLO91}
R. A. M. Klomp, V. G. J. Stoks, and J. J. de
Swart, Phys. Rev. {\bf C44}, R1258 (1991).

\bibitem{STO93}
V. Stoks, R. Timmermans, and J.J. de Swart, Phys. Rev. {\bf C47}, 512 (1993).

\bibitem{ERI95}
T. E. O. Ericson, B. Loiseau, J. Nilsson, N. Olsson, J. Blomgren et al.,
Phys. Rev. Letters {\bf 75},1046 (1995).

\bibitem{FUC90}
N. H. Fuchs, H. Sazdjian and J. Stern, Phys. Lett.  {\bf B238}, 380 (1990).

\bibitem{DOM85}
C. A. Dominguez, Riv. Nuov. Cim.  {\bf 8}, 1 (1985).

\bibitem{OLS96}
J. Blomgren, N. Olsson and {\it et al.,} in preparation.

\bibitem{ERI96}
T. E. O. Ericson, B. Loiseau, J. Blomgren and N. Olsson, $\Pi N$ Newsletter 12, in press.

\bibitem{CHE58}
G.F. Chew, Phys. Rev. {\bf 112}, 1380 (1958).

\bibitem{BIR94}
R. Birsa, F. Bradamante, A. Bressan, S. Dalla Torre-Colautti, et al., Phys. Lett. {\bf B339}, 325
(1994).

\bibitem{BRA94a}
F. Bradamante and A. Martin, Phys. Lett. {\bf B343}, 427 (1995).

\bibitem{BRA94b}
F. Bradamante, A. Bressani, M., M. Lamanna, and A. Martin,  Phys. Lett. {\bf B343}, 431 (1995).

\bibitem{BYS78}
J. Bystricky, F. Lehar, and P. Winternitz, Jour. Phys. {\bf 39}, 1 (1978).

\bibitem{LAF92} P. LaFrance, F. Lehar, B. Loiseau, and P. Winternitz,Helv. Phys.
 Acta {\bf 65}, 611 (1992).

\bibitem{GIB94}
W.R. Gibbs and B. Loiseau, Phys. Rev. {\bf C50}, 2742 (1994).

\bibitem{massdif} The neutron-proton mass difference implies also a small correction to eqs. (3) and
(4) (see ref. 14).  This is neglected, since the effect on the coupling constant is minute 

\bibitem{PIG94}
 M. Pignone, M. Lacombe, B. Loiseau, and
R. Vinh Mau, Phys. Rev. {\bf C50}, 2710 (1994).

\bibitem{TIM95}
R. Timmermans, private communication, see
also R. Timmermans, Th. A. Rijken, and J.J. de Swart, Phys. Rev. {\bf C50},
48 (1994).

\bibitem{HAM80} R.P. Hamilton, T. P. Pun, R. D. Tripp, H. Nickolson,  et al. Phys. Rev. Lett. {\bf
44}, 1179 (1980).

\bibitem{NAK84}  K. Nakamura, T. Fujii, T. Kageyama, F. Sai,  et al., Phys. Rev.
Lett. {\bf53}, 885 (1984).

\bibitem{RON92}
T. R\"onnqvist, H. Cond\'e, N. Olsson, R. Zorro, et al., Phys. Rev. {\bf C45}, R496 (1992).
\end{thebibliography}
\end {document}